\documentclass[manuscript]{aastex}

 \newcommand{\bq}{\begin{equation}} 
 \newcommand{\eq}{\end{equation}}
 \newcommand{\bqn}{\begin{eqnarray}}
 \newcommand{\eqn}{\end{eqnarray}}
 \newcommand{\nb}{\nonumber}
 \newcommand{\lb}{\label}
 \newcommand{\f}{\frac}
 
\newcommand{\PRL}{Phys. Rev. Lett.}

\begin{document}


\title{Detecting quantum gravitational effects of loop quantum cosmology  in the early universe?}

\author{Tao Zhu ${}^{a, b}$, Anzhong Wang ${}^{a, b}$\footnote{The corresponding author}, 
Gerald Cleaver ${}^{c}$, 
Klaus Kirsten ${}^{d}$, Qin Sheng ${}^{d}$, and Qiang Wu ${}^{a}$}

\affil{  ${}^{a}$Institute for Advanced Physics $\&$ Mathematics, Zhejiang University of Technology, Hangzhou, 310032, China\\
 ${}^{ b}$ GCAP-CASPER, Physics Department, Baylor University, Waco, TX 76798-7316, USA\\
  ${}^{c}$ EUCOS-CASPER, Physics Department, Baylor University, Waco, TX 76798-7316, USA\\
 ${}^{d}$ GCAP-CASPER, Mathematics Department, Baylor University, Waco, TX 76798-7328, USA  }

\begin{abstract}
We derive the primordial power spectra and spectral indexes of the density fluctuations and gravitational waves in the framework 
of loop quantum cosmology (LQC) with holonomy and inverse-volume corrections, by using the uniform asymptotic approximation 
method to its third-order, at which the upper error bounds are $\lesssim 0.15\%$, and accurate enough for the current and forthcoming 
cosmological observations. Then, using the Planck, BAO and SN data we obtain the tightest constraints on quantum gravitational effects 
from LQC corrections, and find that such effects could be well within the detection of the current and forthcoming cosmological observations.
\end{abstract}




\section{Introduction}

Quantization of gravity has been one of the main driving forces in physics in the past decades \citep{QGs}, and various  approaches have been pursued, including string/M-Theory  \citep{string},  loop quantum gravity  \citep{LQG}, and more recently the Horava-Lifshitz theory \citep{Horava}. However, it is fair to say that our understanding of it is still  highly limited, and none of the aforementioned approaches is  complete. One of the main reasons is the lack of evidences of quantum gravitational effects, due to  the extreme weakness of gravitational fields.  

This situation has been dramatically changed recently, however, with the arrival of the era of precision cosmology  \citep{QGEs,QGEs1,QGEs2}. In 
particular, cosmic inflation \citep{Guth}, which  is assumed to have taken place during the first moments of time, provides the simplest and 
most elegant mechanism to produce the primordial density perturbations and  gravitational waves.  
The former is responsible for the formations  of the cosmic microwave background (CMB) and the large-scale structure of the universe \citep{DB}.  Current
measurements of CMB   \citep{cosmo,cosmo1,cosmo2} and observations of the large scale distributions of dark matter and galaxies in the universe
\citep{galaxiesS,galaxiesS1,galaxiesS2,galaxiesS3} are in stunning agreement with  it. 
On the other hand, since  inflation is extremely  sensitive to the Planckian physics  \citep{DB,BBCQ,BBCQ1}, it also provides   
 opportunities  to get deep insight into the physics at the energy scales that cannot be reached  by any of  man-made terrestrial  experiments
in the near future. In particular, it  provides a  unique window to explore quantum gravitational  effects from different theories  of quantum gravity, whereby one can
 falsify some of  these theories with observational data that have the uncomprehended accuracy \citep{S4-CMB}, and obtain experimental evidences and valuable 
guidelines for the final construction of   the theory of quantum gravity.

In this Letter, we shall study the quantum gravitational effects of LQC  
  in inflation \citep{LQCs,LQCs1,LQCs2}, and show explicitly that these effects could be 
well within the detection of the current and forthcoming cosmological  experiments \citep{S4-CMB}. 
Such effects  can be studied by introducing appropriate modifications at the level 
of the classical Hamiltonian, very much similar to those studied in solid state physics \citep{LQCs,LQCs1,LQCs2}.  
 It was found that there are mainly two kinds of quantum corrections: 
the holonomy \citep{Mielczarek2008,Mielczarek2009, Mielczarek2010, Grain2010, Li2011, Mielczarek2012, scalar,scalar2,Mielczarek2014}, 
and  inverse-volume  corrections \citep{Bojowald2008,Bojowald2008-1,Bojowald2008-2,Bojowald2008-3,Bojowald2008-4,Bojowald2008-5,Bojowald2011,Bojowald2011a,Bojowald2011b}.  
These corrections modify not only the  linear   
perturbations, but also the space-time background. 

In particular,  for a scalar field $\phi$ with its potential $V(\phi)$,  the holonomy corrections modify 
 the Friedmann and Klein-Gordon equations to the forms, 
 \bqn
 \lb{eq1a}
&&  {\cal{H}}^2 = \frac{8\pi Ga^2\rho_{\phi}}{3}\left(1 - \frac{\rho_{\phi}}{\rho_c}\right),\\
 \lb{eq1b}
&& \phi'' + 2{\cal{H}}\phi' + V_{,\phi} = 0,
 \eqn
 where   $a$ denotes the expansion factor, $ {\cal{H}} \equiv a'/a$,  and a prime denotes the derivative with respect to 
  the conformal time $\eta\;  (\equiv \int{dt/a(t)})$.  
 $\rho_c$ is a constant and characterizes  the  energy scale of the holonomy corrections, with  $\rho_{\phi}
 = {\phi'}^2/(2a^2) + V(\phi)$.
Clearly,  the big bang singularity normally appearing at $\rho_{\phi} = \infty$ now is replaced by a big  bounce occurring at $\rho_{\phi} = \rho_c$. In the infrared (IR) we have
$\rho_{\phi}/ \rho_c \ll 1$, and Eq.(\ref{eq1a})   reduces to that  of general relativity (GR). The evolutions of the anomaly-free cosmological scalar and tensor perturbations 
 are described by the mode function $\mu_k(\eta)$, satisfying the equation   \citep{scalar2,scalar}, 
 \bqn\lb{EoM}
\mu''_k(\eta)+\left(\omega_k^2(\eta)-\frac{z''(\eta)}{z(\eta)}\right)\mu_k(\eta)=0,
\eqn
where  
$\omega_k^2(\eta) = \Omega(\eta)k^2$ with $\Omega(\eta)\equiv1-2\rho/\rho_c$. The background-dependent function $z(\eta)$ is given by
$z_S \; (\equiv  a{\phi'}/{\cal{H}})$ for the scalar perturbations, and  $z_T \; (\equiv a/\sqrt{\Omega})$ for the tensor ones. To the first-order of the slow-roll parameters
and   $\delta_H (\equiv \rho/\rho_c \ll 1)$, the inflationary spectra and spectral indexes with the holonomy corrections have been recently
 obtained,   by further assuming that    the slow-roll parameters and  $\delta_{H}$ are all constants  \citep{Mielczarek2014}.

With the inverse-volume corrections, on the other hand, the Friedmann and Klein-Gordon equations are modified to the forms
\citep{Bojowald2011},
\bqn
\lb{eqa}
&&{\cal{H}}^2 = \frac{8\pi G \alpha}{3}\left(\frac{{\phi'}^2}{2\vartheta} + p V(\phi)\right),\\
&& \phi'' + 2{\cal{H}}\left(1 - \frac{d\ln \vartheta}{d\ln p}\right)\phi' + \vartheta p V_{,\phi} = 0,
\eqn
in which  $ p \equiv a^2$,  $\alpha \equiv 1 + \alpha_0 \delta_{PL} + {\cal{O}}\left(\delta_{Pl}^2\right)$, 
$\vartheta \equiv 1 + \vartheta_0 \delta_{Pl} + {\cal{O}}\left(\delta_{Pl}^2\right)$, and  $\delta_{Pl} \propto a^{-\sigma}$,
 where $\alpha_0, \; \vartheta_0$ and $\sigma$ are   constants [Note that here we use $\vartheta$ instead of $\nu$ adopted   in \citep{Bojowald2011}, and reserve
 $\nu$ for other uses.].
The values of $\alpha_0$ and $ \sigma$  are currently subject to quantization ambiguities, while
the magnitude of $\delta_{Pl} $ is unknown, as so far we have no control  over the details of the underlying full theory of quantum gravity \citep{Bojowald2011}. 
However, when $\sigma$ takes values in the range  $0 < \sigma \le 6$,  the size of $\delta_{Pl} $ does not depend on $\alpha_0$ and $\vartheta_0$, and can be written in the form
 $\delta_{Pl} \equiv (a_{Pl}/a)^{\sigma}$, where $ a_{Pl}$ is another arbitrary constant.
The constant   $\vartheta_0$ is related to $\alpha_0$ and $\sigma$ via the 
consistency relation 
$
\vartheta_0 (\sigma-3)(\sigma+6) - 3\alpha_0(\sigma-6) = 0$.
However, to make the effective theory viable, we shall assume 
$\delta_{Pl}(\eta)   \ll 1$ at any given moment, so we can safely drop off 
all the second- and high-order  terms of $\delta_{Pl}(\eta)$. This assumption also guarantees that the
slow-roll conditions can be imposed, even after the inverse-volume corrections are taken into account. 

With the above assumption, Bojowald and Calcagni (BC)  \citep{Bojowald2011} studied the scalar and 
 tensor perturbations with the inverse-volume corrections, and found that the corresponding mode 
function $\mu_{k}(\eta)$ can be also cast  in the form (\ref{EoM}),  but now with    
\bqn
\lb{equ2}
\omega^{2}_{k}(\eta) =\left(1+2 \alpha_0 \delta_{Pl}(\eta)\right) k^2
\eqn
for tensor, and
\bqn
\omega_k^2(\eta)=\left(1+ 2\beta_0 \delta_{Pl}(\eta)\right)k^2
\eqn
for scalar, where $\beta_0 \equiv {\sigma \vartheta_0} \left({\sigma}+6\right)/36+{\alpha_0}\left(15-{\sigma}\right)/12$.  With such modified dispersion relations, 
BC calculated the corresponding power spectra and spectral indexes to the first-order of the slow-roll parameters, from which, together with Tsujikawa, they 
found \citep{Bojowald2011a,Bojowald2011b} that the LQC effects are distinguishable from these of the noncommutative geometry or string, as the latter manifest themselves in small scales \citep{NCGs,NCGs1,NCGs2}, while the former mainly at large scales. To find explicitly the observational bounds on the inverse-volume quantum corrections,   they considered the CMB likelihood for the potentials $V(\phi) = \lambda_{n} \phi^n$ and $V(\phi) = V_0 e^{-\kappa \lambda\phi}$,  by using the data of WMAP 7yr together with the large-scale structure, the Hubble constant measurement from the Hubble Space Telescope, supernovae type Ia, and big bang nucleosynthesis \citep{cosmo,WMAP7-1,WMAP7-2,WMAP7-3,WMAP7-4}, the most accurate data available to them by then, and obtained various constraints on $\delta(k)$ for different values  of $\sigma$ at the pivots $k_0 = 0.002\; {\mbox{Mp}} c^{-1}$ and $k_0 = 0.05\; {\mbox{Mp}} c^{-1}$
 where $\delta(k) = \alpha_0 \delta_{Pl}(k)$ for $\sigma \not= 3$, and   $\delta(k) =  \vartheta_0 \delta_{Pl}(k)$ for $\sigma = 3$.  
 An interesting feature is that
 the constraints are very sensitive to the choice of the pivots $k_0$, specially when $\sigma$ is large ($\sigma \ge 2$), but insensitive to the forms of the potential $V(\phi)$.

 In this Letter our goals are two-fold: First, we calculate the scalar and tensor power spectra, spectral indexes and the ratio  $r$  to the second-order  of the slow-roll parameters,
 for both of the holonomy and inverse-volume corrections, so that they are  accurate enough to match   with the accuracy  required by the current and forthcoming experiments \citep{S4-CMB}.
This becomes possible, due to  the recent development of the powerful uniform asymptotical approximation method  \citep{Habib-Uniform,ZWCKS1,ZWCKS1-1,ZWCKS1-2,ZWCKS2}, which is
designed specially for  the studies of inflationary models  after quantum gravitational effects are taken into account. Up to 
the third-order approximations in terms of the free parameter ($\lambda^{-1}$) introduced in the method,
which is independent of the slow-roll inflationary parameters mentioned above, the upper error bounds are less than $0.15\%$ \citep{ZWCKS2}. 
Second, we shall use the most recent observational data to obtain  new constraints on $\delta(k_0)$ for the
 power-law potential $V(\phi) = \lambda_n \phi^n$, where $n$ is chosen so that $r \lesssim 0.1$. With such constraints, we shall prove explicitly
  that the quantum gravitational effects from the inverse-volume corrections
 are within the range of the detection of the forthcoming experiments, specially of the Stage IV ones \citep{S4-CMB}.

\section{Inflationary Spectra and Spectral Indexes}


To apply  the uniform asymptotic approximation method, we first rewrite Eq.(\ref{EoM}) to
$
\frac{d^2\mu_k(y)}{dy^2} =\left[\lambda^2\hat g(y)+q(y)\right]\mu_k(y)$,
where $y \equiv - k\eta$, and the parameter $\lambda$ is a large constant to be used to trace the order of approximations. 
The reason to introduce two functions $\hat g(y)$ and $q(y)$, instead of only one,  is to use the extra degree of freedom to
minimize the errors \citep{ZWCKS1}.
For example, with the holonomy corrections, we have 
$ \lambda^2\hat g(y)+q(y)= {z''}/{(k^2z)} - \Omega(\eta)$.
 Then,  minimizing the error control function defined explicitly in \citep{ZWCKS1}, we find that in this case $q(y)$ 
 must be taken as $q(y)=-1/(4y^2)$.
Once $q(y)$ is determined, $\hat g(y)$ is in turn uniquely fixed. Then, the corresponding approximate analytical solution 
will depend on the number and nature (real or complex) of the roots of the equation $\hat g(y) = 0$ \citep{ZWCKS1,ZWCKS1-1,ZWCKS1-2}. 
In the quasi-de Sitter background, it can be shown that   $\hat g(y)$ currently has only one real root. In this case,  the 
general expressions of the mode 
function,  power spectra and spectral indexes up  to the third-order approximations (in terms of $\lambda^{-1}$)  were given 
explicitly  in \citep{ZWCKS2}. Applying them to the case with the holonomy corrections, we find     \citep{ZWCKS4},
 \bqn
 \lb{scalarH}
 \Delta_s^2(k) &= &A_{s}^{\star} \Bigg[ 1-2\left(1+D_{p}^{\star}\right) \epsilon _{\star 1}-D_{p}^{\star} \epsilon _{\star 2}+\delta _{\star H}+\left(2D_{p}^{\star 2}+2D_{p}^{\star}+\frac{\pi^2}{2}-5+\Delta_1^{\star}\right) \epsilon _{\star 1}^2\nb\\
 &&~~~~+\left(\frac{1}{2}D_{p}^{\star 2}+\frac{\pi^2}{8}-1+\frac{\Delta_1^{\star}}{4}\right) \epsilon _{\star 2}^2+\frac{3}{2} \delta _{\star H}^2-D_{p}^{\star}\delta _{\star H} \epsilon _{\star 2}+\left(\frac{\pi ^2}{24}-\frac{1}{2}D_{p}^{\star 2}+\Delta_2^{\star}\right) \epsilon _{\star 2} \epsilon _{\star 3}\nb\\
 &&~~~~+\left(D_{p}^{\star 2}-D_{p}^{\star}+\frac{7\pi ^2}{12}-7+\Delta_1^{\star}+2\Delta_2^{\star}\right) \epsilon _{\star 1} \epsilon _{\star 2}-\left(4 D_{p}^{\star}+6\right) \delta _{\star H} \epsilon _{\star 1}\Bigg],\nb\\
 \Delta_t^2(k) & = & 
A_t^{\star} \Bigg[1+\delta _{\star H}+\frac{3 }{2}\delta _{\star H}^2-2\left(D_{p}^{\star}+1\right) \epsilon _{\star 1}-\left(4 D_{p}^{\star}+6\right) \delta _{\star H} \epsilon _{\star 1}\nb\\
&&~~~~+\left(2D_{p}^{\star 2}+2D_{p}^{\star}+\frac{\pi^2}{2}-5+\Delta_1^{\star}\right) \epsilon _{\star 1}^2+\left(-2D_{p}^{\star 2}-D_{p}^{\star}+\frac{\pi ^2}{12}+2\Delta_2^{\star}\right) \epsilon _{\star 1} \epsilon _{\star 2}\Bigg],\nb\\
 n_s&=&1-2 \epsilon _{\star 1}-\epsilon _{\star 2}+4 \delta _{\star H} \epsilon  _{\star 1}-2 \epsilon _{\star 1}^2-\left(3+2D_{n}^{\star} \right) \epsilon _{\star 1} \epsilon _{\star 2}-D_{n}^{\star}\epsilon _{\star 2} \epsilon _{\star 3},\nb\\
n_t & = &
-2 \epsilon _{\star 1}+4 \delta _{\star H} \epsilon _{\star 1}-2 \epsilon _{\star 1}^2-2\left(D_{n}^{\star}+1\right) \epsilon _{\star 1} \epsilon _{\star 2},\;\;\;\;\;\;
r = 16 \epsilon _{\star 1} (1+D_{p}^{\star}\epsilon _{\star 2}),
\eqn
where $\delta_{H} \equiv \rho_{\phi}/\rho_c \ll 1,\; A_{s}^{\star}\equiv 181 H_{\star }^2/(72 e^3 \pi ^2 \epsilon _{\star 1})$, $A_t^{\star}\equiv 181 H_{\star }^2/(36 e^3 \pi ^2)$, $D_{p}^{\star}=67/181-\ln3$, $D_{n}^{\star}=10/27-\ln3$, $\Delta_1^{\star} =\frac{485296}{98283}-\frac{\pi^2}{2}$, $\Delta_2^{\star} =\frac{9269}{589698}$, and $\star$ denotes quantities evaluated at horizon crossing $a(\eta_\star)H(\eta_\star)=\sqrt{\Omega(\eta_{\star})}k$. $\epsilon_{n}$ denote the slow-roll parameters, defined as $\epsilon_1\equiv -\dot H/H^2$, $\epsilon_{n+1}\equiv\dot \epsilon_n/(H\epsilon_n)\; (n\geq 1)$.
 Note that in the above expressions we have ignored terms at the orders higher than $\mathcal{O}(\epsilon^3,\epsilon^2\delta_H)$. To the first-order, it can be shown that our results are consistent with those presented in \citep{Mielczarek2014}.

In the case with the inverse-volume corrections,   we have $\lambda^2\hat g(y)+q(y)=k^{-2}(z''/z - \omega_k^2(\eta))$, where $\omega_k^2(\eta)$ is given by Eq.(\ref{equ2}), with $z_s(\eta)\equiv a\dot \varphi [1+\frac{1}{2}(\alpha_0-2\vartheta_0)\delta_{Pl}]$ and $z_t(\eta)\equiv a(1-\alpha_0 \delta_{Pl}/2)$, respectively. To minimize the errors,  $q(y)$ must be also chosen as in the last case, and then it can be shown that  $\hat g(y) = 0$ has only one real root, and  as a result,   the general expressions of the mode 
function,  power spectra and spectral indexes  given in   \citep{ZWCKS2} are also applicable to this case, which yield      \citep{ZWCKS4},
 \bqn
 \lb{IVQs}
 \Delta_s^2(k) & =  & A_s\Big[ 1-2\left(1+D_{p}^{\star}\right) \epsilon _{\star 1}-D_{p}^{\star} \epsilon _{\star 2}+\epsilon_{Pl} \left(\frac{3}{2}H_\star\right)^\sigma \left(\mathcal{Q}_{-1}^{\star(s)}\epsilon_{\star1}^{-1}+\mathcal{Q}_{0}^{\star(s)}+\mathcal{Q}_{1}^{\star(s)}\epsilon_{\star2}\epsilon_{\star1}^{-1}\right)\Big],\nb\\
 \Delta_t^2(k) & = & 
A_t\Big[1-2\left(D_{p}^{\star}+1\right) \epsilon_{\star 1}+\epsilon_{Pl} \left(\frac{3}{2}H_\star\right)^\sigma \mathcal{Q}_{0}^{\star(t)}\Big],\nb\\
n_s& = &1-2 \epsilon _{\star 1}-\epsilon _{\star 2}-2 \epsilon _{\star 1}^2-\left(3+2D_{n}^{\star} \right) \epsilon _{\star 1} \epsilon _{\star 2}-D_{n}^{\star}\epsilon _{\star 2} \epsilon _{\star 3}\nb\\
&&+\epsilon_{Pl} \left(\frac{3}{2}H_\star\right)^\sigma \left(\mathcal{K}_{-1}^{\star(s)}\epsilon_{\star1}^{-1}+\mathcal{K}_{0}^{\star(s)}+\mathcal{K}_{1}^{\star(s)}\epsilon_{\star2}\epsilon_{\star1}^{-1}
 \right),\nb\\
n_t & = &
-2 \epsilon _{\star 1}-2 \epsilon _{\star 1}^2-2\left(D_{n}^{\star}+1\right) \epsilon _{\star 1} \epsilon _{\star 2}+\epsilon_{Pl} \left(\frac{3}{2}H_\star\right)^\sigma \mathcal{K}^{\star(t)}_0,\nb\\
r  &=&   16 \epsilon_{\star1} \big[1+D_p\epsilon_{\star2}-\epsilon_{Pl}(\frac{3}{2}H_\star)^{\sigma} \mathcal{Q}^{\star(s)}_{-1} \epsilon_{\star1}^{-1}\big].
\eqn
Note that we parametrize $\delta_{Pl}(\eta)=(a_{Pl}/k)^{\sigma} (-a\eta)^{-\sigma} y^{\sigma}$ with $\epsilon_{Pl}\equiv (a_{Pl}/k)^{\sigma}$, $k \equiv (-a\eta)^{-\sigma}$.
In TABLE I, we list the values of the coefficients $\mathcal{Q}_{-1}^{\star(s)}$, $\mathcal{K}_{-1}^{\star (s)}$, $\mathcal{Q}_0^{\star (t)}$, and $\mathcal{K}_0^{\star (t)}$ for different values of $\sigma$, as they represent  the dominant contributions.  The rest of the terms appearing in the above expressions are subdominant and will not be given here, but they are given explicitly in \citep{ZWCKS4}. When $\sigma=3$, $\mathcal{Q}_{-1}^{\star(s)}$ and $\mathcal{K}_{-1}^{\star (s)}$ vanish, so  one has to consider contributions from $\mathcal{Q}_0^{\star(s)}$ and $\mathcal{K}_0^{\star(s)}$, which  are given by  $\mathcal{Q}_0^{\star(s)}=\frac{513 \pi }{11584}\vartheta_0 $ and $\mathcal{K}^{\star(s)}_0=-\frac{9 \pi}{64} \vartheta_0$.
We emphasize that the modified power spectra and also spectral indices are now explicitly scale-dependent because of $\epsilon_{Pl}\sim k^{-\sigma}$.

Before considering the observational constraints, let us first note that  in \citep{Bojowald2011,Bojowald2011a,Bojowald2011b} the observables $n_s,\; n_t$ and  $r$ were calculated up to the first order of the slow-roll parameters.
Comparing their results with ours, after writing all expressions in terms of the same set of parameters, say, $\epsilon_{V} [\equiv M_{Pl}^2(V_{,\phi}/V)^2/2], \; \eta_{V} [\equiv M_{Pl}^2V_{,\phi\phi}/ V]$, and $
\xi^2_{V} = M_{Pl}^4V_{,\phi} V_{,\phi\phi\phi}/V^2]$,  we find that our results are different from theirs. A closer examination shows that this is mainly due to  the following: (a) In \citep{Bojowald2011} the horizon crossing 
was taken  as $k = {\cal{H}}$. However, due to the quantum gravitational effects, the dispersion relation is modified to the form (\ref{equ2}), so the horizon crossing should be at $\omega_k =  {\cal{H}}$. (b) In \citep{Bojowald2011} the mode function was first obtained at two limits, ${k\gg {\cal{H}}}$ and ${k\ll {\cal{H}}}$, and then   
 matched together at the horizon crossing  where $k \simeq  {\cal{H}}$. 
This may lead  to  huge errors \citep{JM,JM-1}, as neither $\mu_{k\gg {\cal{H}}}$ nor  $\mu_{k\ll {\cal{H}}}$   is a good approximation of the mode function $\mu_{k}$  at the horizon crossing. 
The above arguments can be seen further by considering the exact solution of  
$\mu_{k}$,   
\bqn
\left. \mu_k(\eta)\right|_{\sigma=2} &=& \frac{c_1}{\sqrt{-\eta}} WW\left(-\frac{ia_1}{4\sqrt{a_2}},\frac{\nu}{2},-i \sqrt{a_2} k^2 \eta^2\right),~~~~~
\eqn
for  the case $\sigma = 2$, where $WW(b_1,b_2,z)$ denotes the WhittakerW function,  
$a_1 \equiv 1-m \epsilon_{Pl} \kappa,\;
a_2 \equiv 2\beta_0 \epsilon_{Pl} \kappa$, $m(\eta)$ is the coefficient of $\delta_{Pl}(\eta)$ in the definition $-\f{1}{k^2}\f{z''}{z}=\f{\nu^2-1/4}{y^2}+\f{m}{y^2}\delta_{Pl}$, and $\nu =  3/2 + \epsilon_1 + \epsilon_2/2$ for the scalar perturbations, and $\nu = 3/2 + \epsilon_1$ for the  tensor. Matching it to the Bunch-Davies vacuum solution at
$k \gg {\cal{H}}$, we find that $c_1={e^{-\frac{a_1 \pi}{8 \sqrt{a_2}}}}/({\sqrt{2}k a_2^{1/4}})$. With the above mode function,  the power spectra and spectral indexes can be calculated, and found to be the same as those
given here, but are different from those  of  \citep{Bojowald2011,Bojowald2011a,Bojowald2011b}. For more details, see   \citep{ZWCKS4}.




\section{Detection of Quantum Gravitational Effects}

The contributions to the inflationary spectra and spectral indices from the holonomy corrections are introduced through the parameter $\delta_{\star H}$, 
which are of the order of $10^{-12}$ for typical values of the parameters \citep{Mielczarek2014}. Then,  with  the current and forthcoming observations \citep{S4-CMB},  it is very difficult to detect   such effects. 

On the other hand, for the inverse-volume corrections, let us consider the power-law potential $V(\phi) =  \lambda_n\phi^n$, for which we find that    
$\eta_V = 2(n-1)\epsilon_V/n,\; \xi^2_V= 4(n-1)(n-2)\epsilon_V^2/n^2$, where  $\epsilon_V = M_{Pl}^2 n^2/(2\phi^2)$. Thus, without the  
inverse-volume corrections ($\delta_{Pl} = 0$), we have $n_s = n_s(\epsilon_V)$ and  
$r = r(\epsilon_V)$, and  up to the second-order of   $\epsilon_V$,  the relation \citep{phi2},
\bqn
\lb{CNSTZ}
\Gamma_n (n_s, r) &\equiv& (n_s-1) + \frac{(2+n)r}{8n} \nb\\
&& + \frac{(3n^2 + 18n -4)(n_s-1)^2}{6(n+2)^2}   = 0,
\eqn
holds precisely. The results from Planck 2015    are $n_s = 0.968 \pm 0.006$ and $r_{0.002} < 0.11 (95 \%$ CL) \citep{cosmo,cosmo1,cosmo2}, which yields $n \lesssim 1$.   In the forthcoming experiments, specially the   Stage 
IV ones,  the errors  of the measurements on both $n_s$ and $r$ are  $\le 10^{-3}$ \citep{S4-CMB}, which implies
$\sigma(\Gamma_n) \le 10^{-3}$. On the other hand, when the inverse-volume corrections are taken into account ($\delta_{Pl} \not= 0$), we have $n_s = n_s(\epsilon_V, \epsilon_{Pl})$ and  
$r = r(\epsilon_V,  \epsilon_{Pl})$, and  Eq.(\ref{CNSTZ})  is modified to, 
\bq
\lb{CNSTZb}
\Gamma_n(n_s, r) = \mathcal{F}(\sigma) \frac{\delta(k)}{\epsilon_{V}},
\eq
where $\delta(k)\equiv \alpha_0 \epsilon_{Pl}H^{\sigma}$ and  $\mathcal{F}(\sigma) \simeq \mathcal{O}(1)$ \citep{ZWCKS4}. Clearly, the right-hand side of the above equation represents the quantum gravitational 
effects from the inverse-volume corrections. If it is  equal or greater than 
$ {\cal{O}}(10^{-3})$, these effects shall be well within the detection of the current or  forthcoming experiments. It is interesting to note that the quantum gravitational effects are enhanced by an order $\epsilon_{V}^{-1}$,
which  is absent in \citep{Bojowald2011}.

\begin{figure}
\plottwo{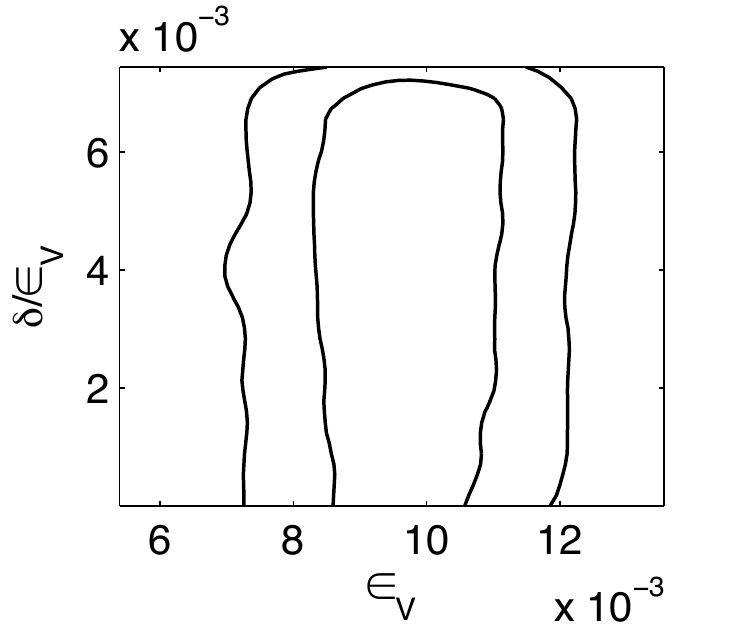}{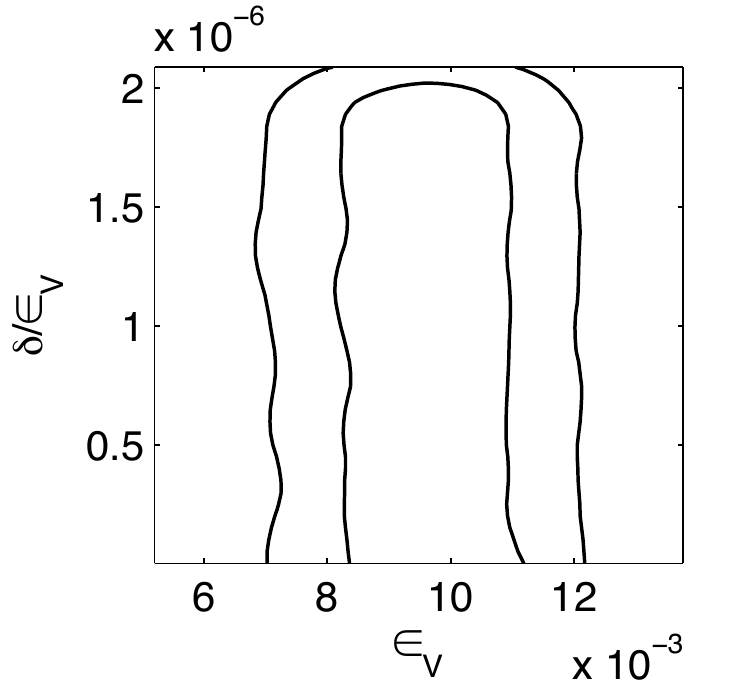}
\caption{Two-dimensional marginalized distribution for the parameters $\delta/\epsilon_V$ and  $\epsilon_V$ at the pivot  $k_0=0.002\mbox{Mpc}^{-1}$ for the power-law potential with $n=1$. (a) The left  panel is for $\sigma=1$, and (b) the right  panel is for $\sigma=2$. The internal and external lines correspond to the confidence levels of  $68\%$ and $95\%$, respectively. \label{fig1}}
\end{figure}

In the following, we run the Cosmological Monte Carlo (CosmoMC) code \citep{COSMOMC} with the Planck \citep{Planck2013}, BAO \citep{BAO2013}, and Supernova Legacy Survey 
\citep{SN} data for the power-law potential with $n=1$, which can be naturally realized in the axion monodromy inflation motived by string/M theory \citep{axion,axion-1}. To compare our results with these acquired in \citep{Bojowald2011b}, we shall carry out our CMB likelihood analysis as closed to theirs as possible. In particular, we assume the flat cold dark matter model with the effective number of neutrinos $N_{eff}=3.046$ and fix the total neutrino mass $\Sigma m_\nu=0.06 eV$. We vary the seven parameters: (i) baryon density parameter, $\Omega_bh^2$, (ii) dark matter density parameter, $\Omega_ch^2$,  (iii) the ratio of the sound horiozn to the angular diameter, $\theta$, (iv) the reionization optical depth $\tau$, (v) $\delta(k_0)/\epsilon_{V}$, (vi) $\epsilon_{V}$, and (vii) $\Delta_s^2(k_0)$. We take the pivot wave number $k_0 = 0.05 \; {\mbox{Mpc}}^{-1}$ used in Planck to constrain $\delta(k_0)$ and $\epsilon_V$. 
In Fig.\ref{fig1}, the constraints on ${\delta}/{\epsilon_{V}}$ and ${\epsilon_{V}}$  are given, respectively,  for   $\sigma=1$ and  $\sigma=2$. In particular, we find that $\delta(k_0)   \lesssim
6.8\times10^{-5}$ ($68\%$ CL) for $\sigma=1$, and $\delta(k_0)   \lesssim 1.9 \times10^{-8}$ ($68\%$ CL) for $\sigma=2$, which are much tighter than those given in  \citep{Bojowald2011b}. The upper bound
for $\delta(k_0)$ decreases dramatically as $\sigma$ increases \citep{Bojowald2011b,ZWCKS4}. However, for any given $\sigma$, the best fitting value of $\epsilon_{V}$  is  about $10^{-2}$, which is rather robust in comparing
with the case without the gravitational quantum effects \citep{Planck2013}. 
It is remarkable to note that, despite the tight constraints on $\delta(k_0)$, because of the $\epsilon_{V}^{-1}$ enhancement of Eq.(\ref{CNSTZb}),  such effects can be well within the range of the
detection of the current and forthcoming cosmological experiments \citep{S4-CMB} for $\sigma \lesssim 1$. Note that small values of $\sigma$ are also favorable theoretically \citep{Bojowald2011}.

\section{Conclusions}


Using the uniform asymptotic approximation method developed recently in \citep{ZWCKS1,ZWCKS1-1,ZWCKS1-2,ZWCKS2}, we have accurately computed the power spectra, spectral indices and the ratio $r$ of the scalar and tensor perturbations of inflation in LQC to the second-order of the slow-roll parameters, after the corrections of the holonomy \citep{Mielczarek2008,Mielczarek2009, Mielczarek2010, Grain2010, scalar,scalar2,Mielczarek2014} and  inverse-volume  \citep{Bojowald2008,Bojowald2008-1,Bojowald2008-2,Bojowald2008-3,Bojowald2008-4, Bojowald2008-5, Bojowald2011,Li2011, Mielczarek2012, Bojowald2011b} are taken into account. The upper error bounds are $\lesssim 0.15\%$, which is accurate enough for the current and forthcoming experiments 
\citep{S4-CMB}. Utilizing the most accurate CMB, BAO and SN data  currently available publicly \citep{Planck2013,BAO2013,SN}, we have carried out the CMB likelihood analysis, and found  constraints 
on $({\delta}(k_0), {\epsilon_{V}})$, the tightest ones obtained so far in the literature. Even with such tight constraints,  the quantum gravitational  effects due to the inverse-volume corrections of LQC can be well within the range of the
detection of the current and forthcoming cosmological experiments \citep{S4-CMB}, provided that $\sigma \lesssim 1$. 

It should be noted that in our studies of the holonomy corrections, the effects of bouncing of the universe are insignificant by implicitly assuming that inflation occurred long 
after the bouncing. This is the same as those considered in \citep{Mielczarek2008,Mielczarek2009, Mielczarek2010, Grain2010, Li2011, Mielczarek2012,scalar,scalar2,Mielczarek2014}. Thus, it is expected that quantum gravitational effects from these corrections are neglectible. However, when the whole process of the bouncing   is properly taken into account, such effects  may not be small at
all   \citep{BG09,BG14}. It would be very interesting to reconsider the observational aspects of these effects, although cautions must be taken,
as Eq.(\ref{eq1a})   was derived only for
small potentials.   Without this condition, there would be additional quantum
corrections which are neither of holonomy nor of inverse-volume type.  
the condition.

\acknowledgments

We thank Shinji Tsujikawa for valuable comments and suggestions.  This work is supported in part by  Ci\^encia Sem Fronteiras, No. A045/2013 CAPES, Brazil (A.W.);
China NSF  Grant, Nos. 11375153 (A.W.), 11047008 (T.Z.), 11105120(T.Z.) and 11205133 (T.Z.); and a URC Award (No. 30330248) from Baylor University (Q.S.)

\clearpage



%
%
%
%



\begin{table}
\begin{center}
\caption{\label{tab:table1} Values of Coefficients $\mathcal{Q}_{-1}^{\star(s)}$, $\mathcal{K}_{-1}^{\star (s)}$, $\mathcal{Q}_0^{\star (t)}$, and $\mathcal{K}_0^{\star (t)}$ for different values of $\sigma$.}
\begin{tabular}{crrrrrrrrrrr}
\tableline\tableline
$\sigma$ & 1 & 2 & 3 &4&5 &6\\
\tableline
$\mathcal{Q}_{-1}^{\star(s)}$ & $\frac{\pi}{6}\alpha_0$ &$\frac{2}{3}\alpha_0$ & 0 &  $\frac{-1616 }{1629}\alpha _0$&$\frac{475}{2896} \pi  \alpha _0$ &$\frac{10512 }{905}\alpha _0$ \\
$\mathcal{K}_{-1}^{\star(s)}$ & $-\frac{\pi}{6}  \alpha _0$ & $-\frac{4}{3}\alpha _0$ & 0 &$\frac{320 \alpha _0}{81} $&$-\frac{125}{144} \pi  \alpha _0 $ & $-\frac{352}{5} \alpha _0 $\\
$\mathcal{Q}_{0}^{\star(t)}$ & $-\frac{725 \pi }{2172} \alpha _0$ & $-\frac{244}{543} \alpha _0$&0 & $\frac{11728}{8145} \alpha _0$& $\frac{8165 \pi  }{5792}\alpha _0$& $\frac{13920 }{1267}\alpha _0$ \\
$\mathcal{K}_{0}^{\star(t)}$ & $\frac{\pi }{3} \alpha _0$ & $\frac{8 }{9}\alpha _0$ & 0& $-\frac{2368}{405}\alpha _0 $& $-\frac{1025 \pi }{144}\alpha _0 $ & $-\frac{6976}{105} \alpha _0$\\
\tableline
\end{tabular}
\end{center}
\end{table}


\begin{thebibliography}{}
\bibitem[Abazajian et al.(2015)]{S4-CMB}Abazajian, K.N., et al. 2015, Astropart. Phys., {\bf 63}, 55
\bibitem[Ade et al.(2013)]{Planck2013} Ade, P. A. R.  (Planck Collaboration) 2014, Astron. Astrophys. {\bf 571}, A16   
\bibitem[Amoros et al.(2014)]{Bojowald2008-5}Amoros, J., de Haro, J., \& Odintsov, S.D. 2014, Phys.\ Rev.\ D, {\bf 89}, 104010
\bibitem[Anderson et al.(2013)]{BAO2013}Anderson, L., et al. 2013, Mon. Not. R. Astron. Soc., {\bf 427}, 3435
\bibitem[Ashoorioon et al.(2011)]{JM-1}Ashoorioon, A., Chialva, D., \& Danielsson, U. 2011,  JCAP, {\bf 06}, 034
\bibitem[Ashtekar \& Singh(2011)]{LQCs1}Ashtekar, A., \& Singh, P. 2011, Class. Quant. Grav., {\bf 28}, 213001


\bibitem[Barrau et al(2014)]{LQCs2}Barrau, A., Cailleteau, T., Grain, J., \&Mielczarek, J. 2014, Class. Quant. Grav., {\bf 31}, 053001
\bibitem[Barrau \& Grain (2014)]{BG14}   Barrau, A. and Grain, J. 2014,  arXiv:14110.1714
\bibitem[Baumann(2009)]{DB}Baumann, D. 2009, arXiv:0907.5424
\bibitem[Becker et al.(2007)]{string} Becker,  K., Becker, M., \& Schwarz, J.H. 2007, {\em String Theory and M-Theory} (Cambridge University Press, Cambridge)
\bibitem[Beutler et al.(2011)]{galaxiesS2}Beutler, F., et al. 2011, Mon. Not. Roy. Astron. Soc., {\bf 416}, 3017
\bibitem[BICEP2/Keck and Planck Collaborations(2015)]{cosmo1}BICEP2/Keck and Planck Collaborations 2015, arXiv:1502.00612
\bibitem[Blake(2011)]{galaxiesS3} Blake, C. 2011, Mon. Not. Roy. Astron. Soc., {\bf 418}, 1725
\bibitem[Bojowald(2005)]{LQCs}Bojowald, M. 2005, Living Rev. Rel., {\bf 8}, 11
\bibitem[Bojowald \& Hossain(2007)]{Bojowald2008}Bojowald, M., \& Hossain, G.M. 2007, Class. Quantum Grav., {\bf 24}, 4801
\bibitem[Bojowald \& Hossain(2008a)]{Bojowald2008-1}Bojowald, M., \& Hossain, G.M. 2008a, Phys. Rev. D, {\bf 77}, 023508
\bibitem[Bojowald \& Hossain(2008b)]{Bojowald2008-2}Bojowald, M., \& Hossain, G.M. 2008b, Phys. Rev. D, {\bf 78}, 063547
\bibitem[Bojowald et al.(2009)]{Bojowald2008-3}Bojowald, M., Hossain, G.M., Kagan, M., \& Shankaranarayanan, S. 2009, Phys. Rev. D, {\bf 79}, 043505
\bibitem[Bojowald et al.(2010)]{Bojowald2008-4}Bojowald, M., Hossain, G.M., Kagan, M., \& Shankaranarayanan, S. 2010, Phys. Rev. D, {\bf 82}, 109903 (E)
\bibitem[Bojowald \& Calcagni(2011)]{Bojowald2011} Bojowald, M., \& Calcagni, G. 2011, JCAP, {\bf 03}, 032
\bibitem[Bojowald et al.(2011a)]{Bojowald2011a} Bojowald, M., Calcagni, G., \& Tsujikawa, S. 2011a, Phys. Rev. Lett., {\bf 107}, 211302
\bibitem[Bojowald et al.(2011b)]{Bojowald2011b} Bojowald, M., Calcagni, G., \& Tsujikawa, S. 2011b, JCAP, {\bf 11}, 046
\bibitem[Brandenberger \& Martin(2013)]{BBCQ1}Brandenberger, R.H., \& Martin,  J. 2013, Class. Quantum. Grav., {\bf 30}, 113001
\bibitem[Burgess et al.(2013)]{BBCQ}Burgess, C.P., Cicoli, M., \& Quevedo, F. 2013, JCAP, {\bf 11}, 003
\bibitem[Burles \& Tytler(1998)]{WMAP7-4}Burles, S., \& Tytler, D. 1998, Astrophys. J., {\bf 499}, 699

\bibitem[Cailleteau et al.(2012a)]{scalar2}Cailleteau, T., Barrau, A., Vidotto, F., \& Grain, J. 2012a, Phys. Rev. D, {\bf 86}, 087301
\bibitem[Cailleteau et al.(2012b)]{scalar}Cailleteau,T., Mielczarek, J., Barrau A., \& Grain, J. 2012b, Class. Quantum Grav. {\bf 29}, 095010
\bibitem[Calcagni \& Tsujikawa(2004)]{NCGs2}Calcagni, G., \&Tsujikawa, S. 2004, Phys. Rev. D, {\bf 70}, 103514
\bibitem[Conley et al.(2011)]{SN}Conley, A., Guy, J., Sullivan, M., Regnault, N., Astier, P., Balland, C., Basa, S., \& Carlberg R.G., et al. 2011, Astrophys. J. Suppl., {\bf 192}, 1
\bibitem[Creminelli et al.(2014)]{phi2}Creminelli, P., Nacir, D.L., Simonovic, M., Trevisan, G., \& Zaldarriaga, M. 2014, \PRL, {\bf 112}, 241303

\bibitem[Eisenstein et al.(2005)]{galaxiesS}Eisenstein, D.J., et al. 2005, APJ., {\bf 633}, 560
\bibitem[Gong et al.(2008)]{COSMOMC}Gong, Y.-G., Wu, Q., \& Wang, A. 2008, Astrophys. J., {\bf 681}, 27 (see also http://cosmologist.info/cosmomc/)
\bibitem[Grain \& Barrau (2009)]{BG09}   Grain, J. and Barrau, A. 2009,  Phys. Rev. Lett. {\bf 102}, 081301 

\bibitem[Grain et al.(2010)]{Grain2010}Grain, J., Barrau, A., Cailleteau, T., \& Mielczarek, J. 2010, Phys. Rev. D, {\bf 82}, 123520
\bibitem[Guth(1981)]{Guth}Guth, A. 1981, Phys. Rev. D, {\bf 23}, 348


 
 \bibitem[Habib et al.(2002)]{Habib-Uniform}Habib, S., Heitmann, K., Jungman, G., \& Molina-Paris, C. 2002, Phys. Rev. Lett., {\bf 89}, 281301
 \bibitem[Ho\v{r}ava(2009)]{Horava}Ho\v{r}ava, P. 2009, Phys. Rev. D, {\bf 79}, 084008
\bibitem[Joras \& Marozzi(2009)]{JM}Joras, S.E., \& Marozzi, G. 2009, Phys. Rev. D, {\bf 79}, 023514

\bibitem[Kiefer(2012)]{QGs} Kiefer, C. 2012, {\em Quantum Gravity}, Third Edition (Oxford Science Publications, Oxford University Press)
\bibitem[Kiefer \& Kramer(2012)]{QGEs}Kiefer, C. \& Kramer, M. 2012, Phys. Rev. Lett., {\bf 108}, 021301
\bibitem[Komatsu et al.(2011)]{cosmo} Komatsu, E. {\em et al.} 2011 (WMAP Collaboration), Astrophys. J. Suppl, {\bf 192}, 18
\bibitem[Kowalski et al.(2008)]{WMAP7-3}Kowalski, M., et al. 2008, Astrophys. J.,  {\bf 686}, 749
\bibitem[Krauss \& Wilczek(2014)]{QGEs1}Krauss, L.M. \& Wilczek, F. 2014, Phys. Rev. D, {\bf 89}, 047501


\bibitem[Li \& Zhu(2011)]{Li2011}Li, Y., \&Zhu,  J.-Y. 2011, Class. Quantum Grav., {\bf 28}, 045007
\bibitem[McAllister et al.(2010)]{axion-1}McAllister, L., Silverstein, E., \& Westphal, A. 2010, Phys. Rev. D, {\bf 82}, 046003
\bibitem[Mielczarek(2008a)]{Mielczarek2008}Mielczarek, J. 2008, JCAP, {\bf 11}, 011
\bibitem[Mielczarek(2009)]{Mielczarek2009}Mielczarek, J. 2009, Phys. Rev. D, {\bf 79}, 123520
\bibitem[Mielczarek et al.(2010)]{Mielczarek2010}Mielczarek, J., Cailleteau, T., Grain, J., \& Barrau, A. 2010, Phys. Rev. D,  {\bf 81}, 104049
\bibitem[Mielczarek et al.(2012)]{Mielczarek2012} Mielczarek, J., Cailleteau, T., Barrau, A., \& Grain, J. 2012, Class. Quantum Grav., {\bf 29}, 085009
\bibitem[Mielczarek(2014)]{Mielczarek2014}Mielczarek, J. 2014, JCAP, {\bf 03}, 048
\bibitem[Piao et al.(2004)]{NCGs1}Piao, Y.S., et al. 2004, Class. Quantum Grav., {\bf 21}, 4455
\bibitem[Planck Collaboration(2013)]{Planck Collaboration}Planck Collaboration 2013, Astron. Astrophys., {\bf 571}, A16
\bibitem[Planck Collaboration(2015)]{cosmo2}Planck Collaboration 2015, arXiv:1502.01582
\bibitem[Reid et al.(2010)]{WMAP7-1}Reid, B.A., et al. 2010, Mon. Not. R. Astron. Soc., {\bf 404}, 60
\bibitem[Riess et al.(2009)]{WMAP7-2}Riess, A.G., et al. 2009, Astrophys. J., {\bf 699}, 539
\bibitem[Rovelli \& Vidotto(2015)]{LQG} Rovelli, C. \& Vidotto, F. 2015, {\em Covariant Loop Quantum gravity} (Cambridge University Press, Cambridge)
\bibitem[Silverstein \& Westphal(2008)]{axion} Silverstein E., \& Westphal, A. 2008, Phys. Rev. D, {\bf 78}, 106003
\bibitem[Tegmark et al.(2006)]{galaxiesS1} Tegmark, M., et al. 2006, Phys. Rev. D, {\bf 74}, 123507
\bibitem[Tsujikawa et al.(2003)]{NCGs}Tsujikawa, S., et al. 2003, Phys. Lett. B, {\bf 574}, 141
\bibitem[Woodard(2014)]{QGEs2}Woodard, R.P. 2014, arXiv:1407.4748

\bibitem[Zhu et al.(2014a)]{ZWCKS1} Zhu, T., Wang, A., Cleaver, G., Kirsten, K., \& Sheng, Q. 2014a, Int. J. Mod. Phys. A, {\bf 29}, 1450142
\bibitem[Zhu et al.(2014b)]{ZWCKS1-1}Zhu, T., Wang, A., Cleaver, G., Kirsten, K., \& Sheng, Q. 2014b, Phys. Rev. D, {\bf 89}, 043507
\bibitem[Zhu et al.(2014c)]{ZWCKS1-2}Zhu, T., Wang, A., Cleaver, G., Kirsten, K., \& Sheng, Q. 2014c, Phys. Rev. D, {\bf 90}, 103517
\bibitem[Zhu et al.(2014d)]{ZWCKS2}Zhu, T., Wang, A., Cleaver, G., Kirsten, K., \& Sheng, Q. 2014d, Phys. Rev. D, {\bf 90}, 063503
 \bibitem[Zhu et al.(2015)]{ZWCKS4}Zhu, T., Wang, A., Cleaver, G., Kirsten, K., \& Sheng, Q. 2015, in preparation (2015).

\end{thebibliography}
\end{document}